\documentclass[10pt, twocolumn, pre, aps, superscriptaddress, showpacs]{revtex4-1}
\usepackage{amsmath, graphicx, subfigure, tikz}
\begin{document}

\title{Global Ashkin-Teller Phase Diagrams in Two and Three Dimensions:\\
Multicritical Bifurcation versus Double Tricriticality - Endpoint}
\author{Ibrahim Ke\c{c}o\u{g}lu}
    \affiliation{Department of Physics, Bo\u{g}azi\c{c}i University, Bebek, Istanbul 34342, Turkey}
    \affiliation{Department of Mathematics, Bo\u{g}azi\c{c}i University, Bebek, Istanbul 34342, Turkey}
\author{A. Nihat Berker}
    \affiliation{Faculty of Engineering and Natural Sciences, Kadir Has University, Cibali, Istanbul 34083, Turkey}
    \affiliation{T\"UBITAK Research Institute for Fundamental Sciences, Gebze, Kocaeli 41470, Turkey}
    \affiliation{Department of Physics, Massachusetts Institute of Technology, Cambridge, Massachusetts 02139, USA}

\begin{abstract}
The global phase diagrams of the Askin-Teller model are calculated in $d=2$ and 3 by renormalization-group theory that is exact on the hierarchical lattice and approximate on the recently improved Migdal-Kadanoff procedure.  Three different ordered phases occur in the dimensionally distinct phase diagrams that reflect three-fold order-parameter permutation symmetry, a closed symmetry line, and a quasi-disorder line.  First- and second-order phase boundaries are obtained.  In $d=2$, second-order phase transitions meeting at a bifurcation point are seen.  In $d=3$, first- and second-order phase transitions are separated by tricritical and critical endpoints.
\end{abstract}
\maketitle

\section{Introduction: The Ashkin-Teller Model and Its Special Lines}

The Ashkin-Teller model \cite{AT,Kadanoff0} is, literally, a doubled Ising model, and introduces a variety of new ordering phenomena.  The simplest Ashkin-Teller model is defined by the Hamiltonian
\begin{equation}
- \beta {\cal H} = \sum_{\left<ij\right>} \, [J (s_i s_j + t_i t_j) + M s_is_jt_it_j ],
\end{equation}
where $\beta=1/k_{B}T$, at each site $i$ there are two Ising spins $s_i=\pm 1, t_i=\pm 1$, and the sum is over all interacting quadruples of spins on nearest-neighbor pairs of sites.  We consider $J\geq 0$ with no loss of generality, since $J \rightarrow -J$ by redefining the spin directions on one sublattice.

\begin{figure}[ht!]
\centering
\includegraphics[scale=0.40]{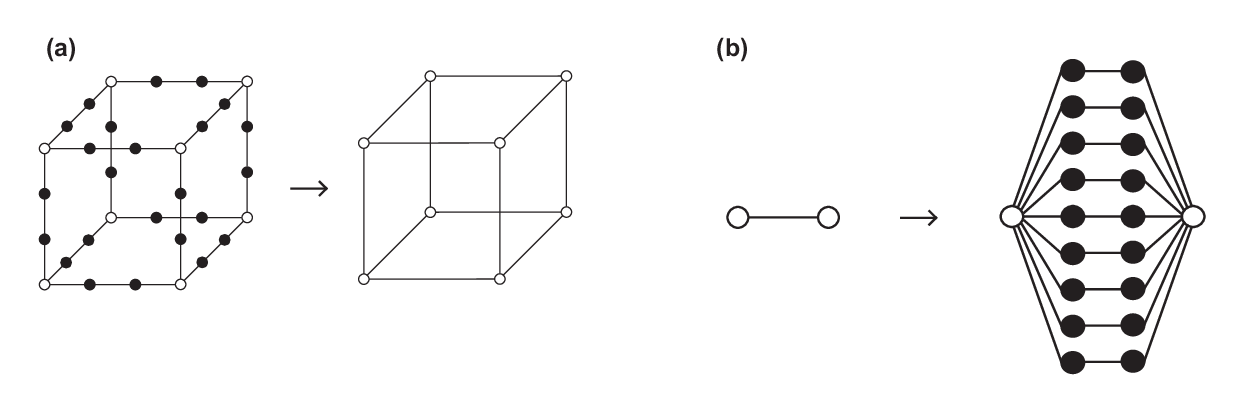}
\caption{(a) The Migdal-Kadanoff approximate renormalization-group transformation on the cubic lattice. Bonds are removed from the cubic lattice to make the renormalization-group transformation doable.  The removed bonds are compensated by adding their effect to the decimated remaining bonds.  (b) A hierarchical model is constructed by self-imbedding a graph into each of its bonds, \textit{ad infinitum}.\cite{BerkerOstlund}  The exact renormalization-group solution proceeds in the reverse direction, by summing over the internal spins shown with the dark circles.  Here is the most used, so called "diamond" hierarchical lattice \cite{BerkerOstlund,Kaufman1,Kaufman2,BerkerMcKay}.  The length-rescaling factor $b$ is the number of bonds in the shortest path between the external spins shown with the open circles, $b=3$ in this case. The volume rescaling factor $b^d$ is the number of bonds replaced by a single bond, $b^d=27$ in this case, so that $d=3$.}
\end{figure}

At each site i, another spin can be defined, $\sigma_i \equiv s_i t_i$, so that the product of any two of $(s_i,t_i,\sigma_i)$ gives the third one.  Thus, in the phase diagram, the $J=M$ line is a symmetry line of the model.  Furthermore, on this $J=M$ line, the Ashkin-Teller model, Eq. (1), reduces to the 4-state Potts model,
\begin{equation}
- \beta {\cal H} = \sum_{\left<ij\right>} \, 3J\, \delta(s_i,s_j)\delta(t_i,t_j)
\end{equation}
where the delta function $\delta(s_i,s_j)=1(0)$ for $s_i=s_j (s_i \neq s_j)$.
In fact, as we shall see below, the Ashkin-Teller multicritical bifurcation point resulting from our calculation occurs exactly on this line and, along the $J=M$ direction, is the 4-state Potts phase transition, which is multicritical (critical-tricritical merged) in $d=2$ and first-order in $d=3$ \cite{spinS7,AndelmanPotts1,AndelmanPotts2,Devre}, as also reflected in our present results.

Furthermore, the energy between interacting sites, as specified in Eq. (1), is in general $2J+M$ with multiplicity 4, $-M$ with multiplicity 8, $-2J+M$ with multiplicity 4.  On the line $J=-M$, this reduces to the site pair having energy $J$ with multiplicity 12 and $-3J$ with multiplicity 4.  The former, highly degenerate states are the ground states, so that the line $J=-M$ is a near-disorder line.  In fact, in both $d=2$ and 3, in our calculated phase diagrams, the entirety of this line is in the disordered state (narrowly in $d=3$, where indeed order should be more dominant due to higher connectivity), down to zero temperature, i.e., up to infinite coupling $J\rightarrow \infty$.

\begin{figure*}[ht!]
\centering
\includegraphics[scale=0.6]{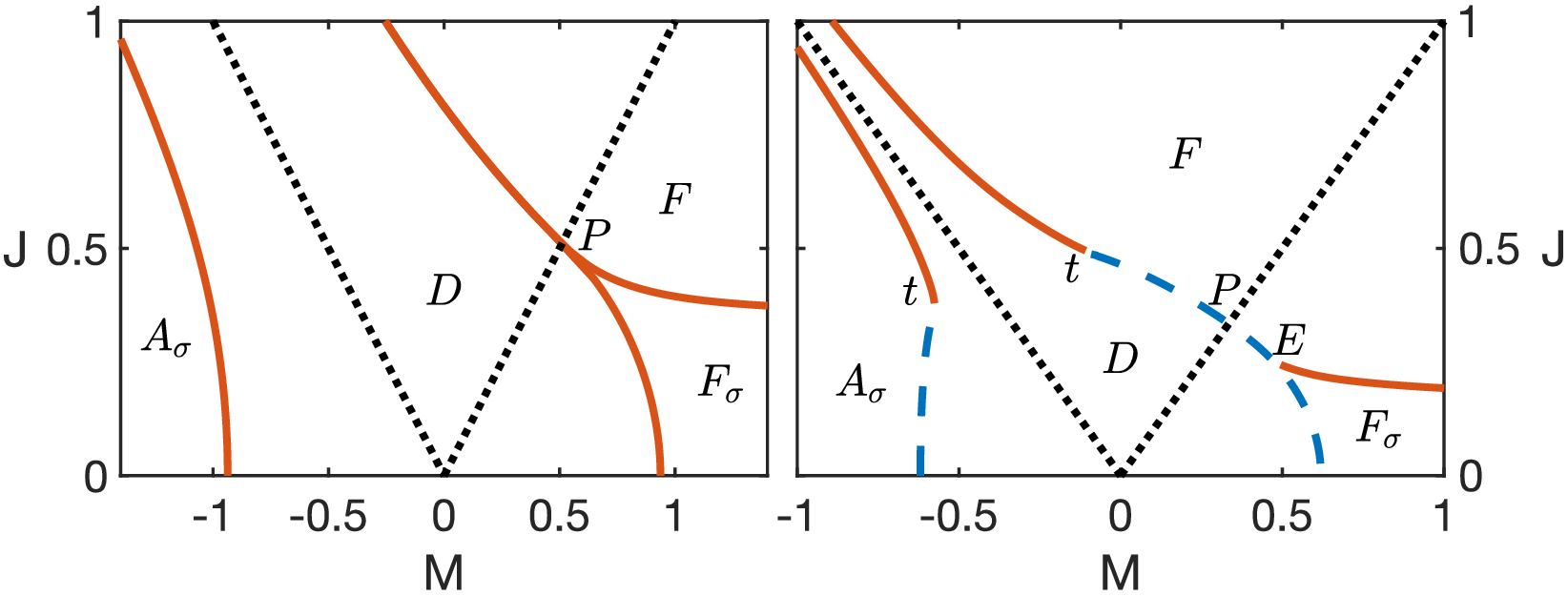}
\caption{Calculated phase diagrams of the Ashkin-Teller model. The $st$-ferromagnetic $(F)$, $\sigma$-ferromagnetic $(F_\sigma)$, $\sigma$-antiferromagnetic $(A_\sigma)$ ordered phases and disordered $(D)$ phase are seen in both dimensions. The dotted lines are the symmetry $(M>0)$ and near-disorder $(M<0)$ lines (see Sec.I). The 4-state Potts phase transition occurs on the symmetry line at point $P$. Left: In the $d=2$, only second-order phase transitions are seen and are marked by full lines.  The bifurcation point, where three critical lines meet and which is also the 4-state Potts transition (second order), occurs exactly on the symmetry line. Right: In the $d=3$, first- and second-order phase transitions are seen and are respectively marked by dashed and full lines. Two tricritical points $t$ and a critical endpoint $E$ occur.  The 4-state Potts phase transition, which is first order here, occurs at point $P$.}
\end{figure*}

\begin{figure}[ht!]
\centering
\includegraphics[scale=0.6]{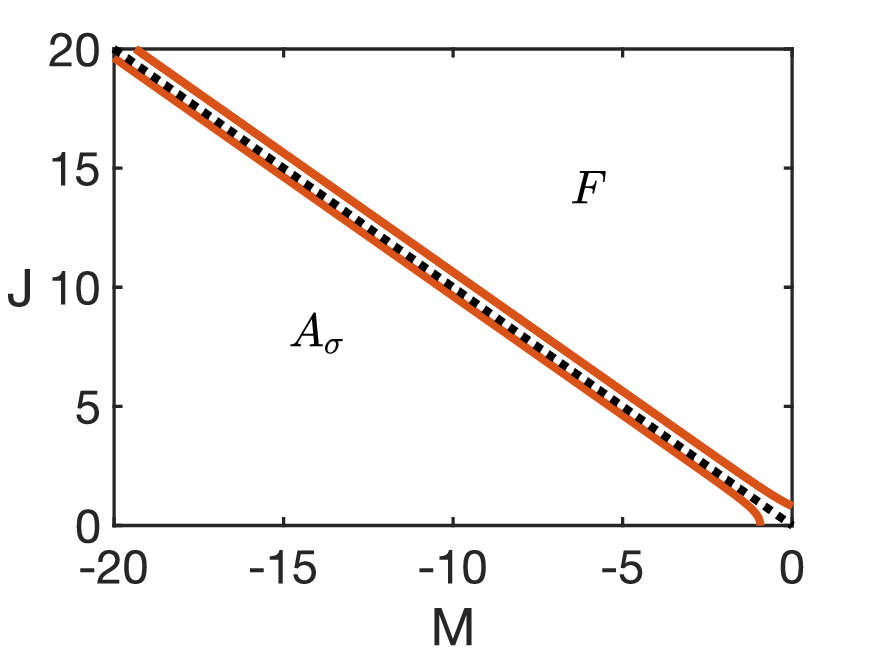}
\caption{The quasi-disorder line (dotted) persists, with a narrow band of disordered phase, to all strong couplings.  Shown here is $d=2$.  The full lines are the calculated second-order phase transitions.}
\end{figure}

We have solved this model in spatial dimensions $d=2$ and 3, with renormalization-group theory exactly on hierarchical lattices and, equivalently, by the recently improved Migdal-Kadanoff approximation on square and cubic lattices. In the space of $(J,M)$ of the phase diagrams, $(s_it_i)$-ferromagnetic, $\sigma$-ferromagnetic $(F_\sigma)$, $\sigma$-antiferromagnetic $(A_\sigma)$ ordered phases and a disordered phase occur, separated by second-order phase transitions in $d=2$ and first- and second-order phase transitions in $d=3$.  The two phase diagrams thus respectively exhibit a distinctive bifurcative multicritical point where three critical lines meet on the symmetry line in $(d=2)$ and first-order phase transitions ushered by two tricritical points and supporting a critical endpoint in $(d=3)$.

\section{Method: Exact Hierarchical and Improved Migdal-Kadanoff}

Hierarchical models \cite{BerkerOstlund,Kaufman1,Kaufman2,BerkerMcKay} are exactly solvable microscopic models that are widely used.\cite{Clark,Kotorowicz,ZhangQiao,Jiang,Chio,Myshlyavtsev,Derevyagin,Shrock,Monthus,Sariyer,MusicBrain}  The construction of a hierarchical model is illustrated in Fig. 1(b) \cite{BerkerOstlund}.  The exact renormalization-group solution of the hierarchical model proceeds in the direction reverse to its construction (Fig. 1), by summing over the internal spins shown in the figures with the dark circles.  In addition to phase transitions \cite{BerkerOstlund,Kaufman1,Kaufman2,BerkerMcKay} and spin-glass chaos \cite{McKayChaos,McKayChaos2,McKayChaos4}, most recently hierarchical models have been developed for the classification and clustering of complex phenomena, including multicultural music and brain electroencephalogram signals \cite{MusicBrain}.

Preceding hierarchical models, the Migdal-Kadanoff approximation is a physically inspired, widely applicable and easy to implement, most used and most successful renormalization-group approximation.\cite{Migdal,Kadanoff}  For example, the physical cubic lattice (Fig. 1a) cannot be directly renormalization-group transformed.  Thus, some of the bonds are removed to make this transformation doable and the transformation is achieved by decimation, namely summing, in the partition function, over the intermediate spins (in black in Fig. 1a).  However, the erstwhile removal of some bonds have weakened the connectivity of the system.  Thus, the effect of every bond removed is added to the decimated bonds. In addition to the application to physical systems such as surface systems \cite{BOP} and high-temperature superconductivity \cite{highTc}, the Migdal-Kadanoff approximation gives the correct lower-critical dimensions and ordering behaviors for all spin systems, including complex quenched-random systems.

As seen in Figs. 1, the Migdal-Kadanoff approximate solution is algebraically identical with the exact solution of a hierarchical model, which makes the Migdal-Kadanoff approximation a physically realizable approximation, as is used in turbulence \cite{Kraichnan}, polymers \cite{Flory,Kaufman}, electronic systems \cite{Lloyd}, and therefore a robust approximation (e.g., it will never yield a negative entropy).

The transformation is best implemented algebraically by writing the transfer matrix between two neighboring sites, namely
\begin{multline}
\textbf{T}_{ij} \equiv e^{-\beta {\cal H}_{ij}} = e^{J (s_i s_j + t_i t_j) + M s_is_jt_it_j} = \\
\left(
\begin{array}{cccc}
e^{2J+M} & e^{-M} & e^{-M} & e^{-2J+M} \\
e^{-M} & e^{2J+M} & e^{-2J+M} & e^{-M} \\
e^{-M} & e^{-2J+M} & e^{2J+M} & e^{-M} \\
e^{-2J+M} & e^{-M} & e^{-M} & e^{2J+M}\end{array} \right),
\end{multline}
where the consecutive states are $(s_i,t_i) = (+1,+1),(+1,-1),(-1,+1),(-1,-1)$.  The decimation step consists in matrix-multiplying $b$ transfer matrices.  The bond-moving step consists in taking, after decimation, the power of $b^{d-1}$ of each element in the transfer matrix.  Here $b$ is the length-rescaling factor of the renormalization-group transformation.  A recent important improvement \cite{Devre} has been to include the local disorder state into the two-site transfer matrix of Eq.(3).  Inside an ordered region of a given $(s,t)$ value, a disordered site does not significantly contribute to the energy in Eq.(1), but has a multiplicity of $4-1=3$, the substraction being because the disordered site cannot be in the $(s,t)$ state of its surrounding ordered region.  This is equivalent to the exponential of an on-site energy and, with no approximation, is shared on the transfer matrices of the $2d$ incoming bonds.  The transfer matrix becomes
\begin{multline}
\textbf{T}_{ij} =
\left(
\begin{array}{ccccc}
e^{2J+M} & e^{-M} & e^{-M} & e^{-2J+M} & 3^{1/2d} \\
e^{-M} & e^{2J+M} & e^{-2J+M} & e^{-M} & 3^{1/2d} \\
e^{-M} & e^{-2J+M} & e^{2J+M} & e^{-M} & 3^{1/2d} \\
e^{-2J+M} & e^{-M} & e^{-M} & e^{2J+M} & 3^{1/2d} \\
3^{1/2d} & 3^{1/2d} & 3^{1/2d} & 3^{1/2d} &3^{1/d}\end{array} \right).
\end{multline}
The inclusion of the disorder state ensures the possible occurrence of temperature-driven first-order phase transitions \cite{spinS7,AndelmanPotts1,AndelmanPotts2}.  This improved renormalization-group transformation has yielded the correct changeover number of states of $q_c=4$ and $q_c=2$ in the $q$-state Potts models in two and three dimensions, respectively.\cite{Devre}

\begin{table*}
\begin{tabular}{c c c c c c c}
\hline

\vline & & & Sinks of the Ordered Phases of the Ashkin-Teller Model in d=2 and 3 & &  &\vline  \\
\hline
\vline & [1,0,0,0] &\vline  & [1,0,0,1]  &\vline  & [0,1,1,0] &\vline   \\
\hline
\vline & $st$ Ferromagnetic  &\vline & $\sigma$ Ferromagnetic  &\vline & $\sigma$ Antiferromagnetic &\vline \\
\hline

\end{tabular}
\caption{Under repeated renormalization-group transformations, the phase diagram points of the ordered phases of the Ashkin-Teller model flow to the sinks shown on this Table.  Only the top row of the sink transfer matrix is shown here.  The subsequent rows can be deduced by symmetry.  The last row and last column of the transfer matrix, corresponding to the effective vacancies, have all elements zero at the ordered sinks and are not given here.}
\end{table*}

\section{Results: Phase Diagrams Subtended by Multicritical Bifurcation, Double Tricriticality and Endpoint}

Under repeated renormalization-group transformations, the phase diagram points of the ordered phases flow to the sinks shown in Table I.  The sink values of the transfer matrix elements epitomize the whole basin of attraction of the completely stable fixed point that is the sink.  The disordered phase has two sinks, one sink with the lower-right $5 \times 5$ element of the transfer matrix equal to 1 and the rest zero, another sink with all elements in the upper-left $4 \times 4$ block equal to one and the rest zero.  Analysis at the unstable fixed points attracting the phase boundaries give the order of the phase transition.\cite{BOP}

The calculated phase diagrams of the Ashkin-Teller model are shown in Fig. 2. The $st$-ferromagnetic $(F)$, $\sigma$-ferromagnetic $(F_\sigma)$, $\sigma$-antiferromagnetic $(A_\sigma)$ ordered phases and disordered $(D)$ phase are seen in both dimensions. The dotted lines are the symmetry $(M>0)$ and near-disorder $(M<0)$ lines (see Sec.I). The 4-state Potts phase transition occurs on the symmetry line at point $P$.

In the $d=2$, only second-order phase transitions are seen.  The bifurcation point, where three critical lines meet and which is also the 4-state Potts transition (second order in here), occurs exactly on the symmetry line. In the $d=3$, first- and second-order phase transitions are seen and are respectively marked by dashed and full lines. Two tricritical points $t$ and a critical endpoint $E$ occur.  The 4-state Potts phase transition, which is first-order here, occurs at point $P$.

\section{Conclusion}

We have solved the Ashkin-Teller model exactly on $d=2$ and 3 hierarchical lattices, obtaining dimensionally distinctly first- and second-order phase transitions, but a similar phase-diagram topology in both dimensions.  We identify a symmetry line on which the 4-state Potts transition occurs, as second order in $d=2$ and first order in $d=3$.  We also identify a near-disorder (near-degeneracy) line, around which the disordered phase persists, be it narrowly, to infinite couplings.  A multicritical bifurcation point $(d=2)$ and double tricritical points and a critical endpoint $(d=3)$ are calculated.

\begin{acknowledgments}
Support by the Academy of Sciences of Turkey (T\"UBA) is gratefully acknowledged.
\end{acknowledgments}


\begin{references}

\bibitem{AT} J. Ashkin and E. Teller, Statistics of Two-Dimensional Lattices with Four Components, Phys. Rev. {\bf 64}, 178 (1943).
\bibitem{Kadanoff0} R. V. Ditzian, J. R. Banavar, G. S. Grest, and L. P. Kadanoff, “Phase Diagram for the Ashkin-Teller Model in Three Dimensions”, Phys. Rev. B {\bf 22}, 2542 (1980).

\bibitem{spinS7} B. Nienhuis, A.N. Berker, E.K. Riedel, and M. Schick, First- and Second-Order Phase Transitions in Potts Models: Renormalization-Group Solution, Phys. Rev. Lett. {\bf 43}, 737 (1979).
\bibitem{AndelmanPotts1} D. Andelman and A. N. Berker, q-State Potts Models in d-Dimensions: Migdal-Kadanoff Approximation, J. Phys. A {\bf 14}, L91 (1981).
\bibitem{AndelmanPotts2} A. N. Berker and D. Andelman, 1st Order and 2nd Order Phase Transitions in Potts Models - Competing Mechanisms, J. Applied Phys. {\bf 53}, 7923 (1982).
\bibitem{Devre} H. Y. Devre and A. N. Berker, First-order to second-order phase transition changeover and latent heats of q-state Potts models in d=2,3 from a simple Migdal-Kadanoff adaptation, Phys. Rev. E {\bf 105}, 054124 (2022).

\bibitem{BerkerOstlund} A. N. Berker and S. Ostlund, Renormalisation-Group Calculations of Finite Systems: Order Parameter and Specific Heat for Epitaxial Ordering, J. Phys. C {\bf 12}, 4961 (1979).
\bibitem{Kaufman1} R. B. Griffiths and M. Kaufman, Spin Systems on Hierarchical Lattices: Introduction and Thermodynamic Limit, Phys. Rev. B {\bf 26}, 5022R (1982).
\bibitem{Kaufman2} M. Kaufman and R. B. Griffiths, Spin Systems on Hierarchical Lattices: 2. Some Examples of Soluble Models, Phys. Rev. B {\bf 30}, 244 (1984).
\bibitem{BerkerMcKay} A. N. Berker and S. R. McKay, Hierarchical Models and Chaotic Spin Glasses, J. Stat. Phys. {\bf 36}, 787 (1984).

\bibitem{Clark} J. Clark and C. Lochridge, Weak-Disorder Limit for Directed Polymers on Critical Hierarchical Graphs with Vertex Disorder, Stochastic Processes and Their Applications {\bf 158}, 75 (2023).
\bibitem{Kotorowicz} M. Kotorowicz and Y. Kozitsky, Phase Transitions in the Ising Model on a Hierarchical Random Graph Based on the Triangle, J. Phys. A, {\bf 55}, 405002 (2022).
\bibitem{ZhangQiao} P.-P. Zhang, Z.-Y. Gao, Y.-L. Xu, C.-Y. Wangmand X.-M. Kong, Phase Diagrams, Quantum Correlations and Critical Phenomena of Antiferromagnetic Heisenberg Model on Diamond-Type Hierarchical Lattices, Quantum Science and Technology {\bf 7}, 025024 (2022).
\bibitem{Jiang} K. Jiang, J. Qiao, and Y. Lan, Chaotic Renormalization Flow in the Potts model induced by long-range competition, Phys. Rev. E {\bf 103}, 062117 (2021).
\bibitem{Chio} I. Chio and R. K. W. Roeder, Chromatic Zeros on Hierarchical Lattices and Equidistribution on Parameter Space, Annales Inst. Henri Poincar\'{e} D {\bf 8}, 491 (2021).
\bibitem{Myshlyavtsev} A. V. Myshlyavtsev, M. D. Myshlyavtseva, and S. S. Akimenko, Classical Lattice Models with Single-Node Interactions on Hierarchical Lattices: The Two-Layer Ising Model, Physica A {\bf558}, 124919 (2020).
\bibitem{Derevyagin} M. Derevyagin, G. V. Dunne, G. Mograby, and A. Teplyaev, Perfect Quantum State Transfer on Diamond Fractal Graphs, Quantum Information Processing, {\bf19}, 328 (2020).
\bibitem{Shrock} S.-C. Chang, R. K. W. Roeder, and R. Shrock, q-Plane Zeros of the Potts Partition Function on Diamond Hierarchical Graphs, J. Math. Phys. {\bf61}, 073301 (2020).
\bibitem{Monthus} C. Monthus, Real-Space Renormalization for Disordered Systems at the Level of Large Deviations, J. Stat. Mech. - Theory and Experiment, 013301 (2020).
\bibitem{Sariyer} O. S. Sar{\i}yer, Two-Dimensional Quantum-Spin-1/2 XXZ Magnet in Zero Magnetic Field: Global Thermodynamics from Renormalisation Group Theory, Philos. Mag. {\bf 99}, 1787 (2019).

\bibitem{MusicBrain}  E. C. Artun, I. Keçoğlu, A. Türkoğlu, and A. N. Berker, Multifractal Spin-Glass Chaos Projection and Interrelation of Multicultural Music and Brain Signals, Chaos, Solitons, Fractals {\bf 167}, 113005 (2023).

\bibitem{McKayChaos} S. R. McKay, A. N. Berker, and S. Kirkpatrick, Spin-Glass Behavior in Frustrated Ising Models with Chaotic Renormalization-Group Trajectories, Phys. Rev. Lett. {\bf 48}, 767 (1982).
\bibitem{McKayChaos2} S. R. McKay, A. N. Berker, and S. Kirkpatrick, Amorphously Packed, Frustrated Hierarchical Models: Chaotic Rescaling and Spin-Glass Behavior, J. Appl. Phys. {\bf 53}, 7974 (1982).
\bibitem{McKayChaos4} S. R. McKay and A. N. Berker, J. Appl. Phys., Chaotic Spin Glasses: An Upper Critical Dimension, J. Appl. Phys. {\bf 55}, 1646 (1984).

\bibitem{Migdal} A. A. Migdal, Phase transitions in gauge and spin lattice systems, Zh. Eksp. Teor. Fiz. {\bf69}, 1457 (1975) [Sov. Phys. JETP {\bf42}, 743 (1976)].
\bibitem{Kadanoff} L. P. Kadanoff, Notes on Migdal's recursion formulas, Ann. Phys. (N.Y.) {\bf100}, 359 (1976).

\bibitem{BOP} A. N. Berker, S. Ostlund, and F. A. Putnam, Renormalization-Group Treatment of a Potts Lattice Gas for Krypton Adsorbed onto Graphite, Phys. Rev. B {\bf 17}, 3650 (1978).
\bibitem{highTc} M. Hinczewski and A.N. Berker, Finite-Temperature Phase Diagram of Nonmagnetic Impurities in High-Temperature Superconductors
   using a d=3 tJ Model with Quenched Disorder, Phys. Rev. B {\bf 78}, 064507 (2008).

\bibitem{Kraichnan} R. H. Kraichnan, Dynamics of Nonlinear Stochastic Systems, J. Math. Phys. {\bf2}, 124 (1961).
\bibitem{Flory} P. J. Flory, Principles of Polymer Chemistry (Cornell University Press: Ithaca, NY, USA, 1986).
\bibitem{Kaufman} M. Kaufman, Entropy Driven Phase Transition in Polymer Gels: Mean Field Theory, Entropy {\bf20}, 501 (2018).
\bibitem{Lloyd} P. Lloyd and J. Oglesby, Analytic Approximations for Disordered Systems, J. Phys. C: Solid St. Phys. {\bf9}, 4383 (1976).

\end{references}
\end{document}